\documentstyle[referee,epsfig]{mn}
\newcommand{\beq}{\begin{equation}}
\newcommand{\eeq}{\end{equation}}
\newcommand{\beqar}{\begin{eqnarray}}
\newcommand{\eeqar}{\end{eqnarray}}
\newcommand{\beqars}{\begin{eqnarray*}}
\newcommand{\eeqars}{\end{eqnarray*}}
\newcommand{\bc}{\begin{center}}
\newcommand{\ec}{\end{center}}
\newcommand{\ben}{\begin{enumerate}}
\newcommand{\een}{\end{enumerate}}
\newcommand{\bit}{\begin{itemize}}
\newcommand{\eit}{\end{itemize}}
\def \cLa{{\cal L}_1^{\pm 1}}
\def \cLb{{\cal L}_2^{\pm 1}}
\def \cLc{{\cal L}_3^{\pm 1}}

\newcommand{\I}{\mbox{\rm i}}

\def \veps{\varepsilon}
\def \d{\partial}
\def \lam{\lambda}

\def\f{{\bf f}-}

\def\r{{\bf r}-}
\def\g{{\bf g}-}
\def\1{{\bf g$_1$}-}
\def\2{{\bf g$_2$}-}

\def\lappreq{\! \stackrel{\scriptscriptstyle <}{\scriptscriptstyle
\sim}\!}

\renewcommand{\(}{\left(}
\renewcommand{\)}{\right)}

\title[Rotational effects on the oscillation frequencies of newly born 
proto-neutron stars]
{Rotational effects on the oscillation frequencies of newly born
proto-neutron stars}
\author[V. Ferrari, L.Gualtieri, J.A. Pons, A. Stavridis]
{V. Ferrari$^1$, L. Gualtieri$^1$, J.A. Pons$^{2}$, A. Stavridis$^1$
\\ $^1$ Dipartimento
di Fisica ``G.Marconi", Universit\` a di Roma ``La Sapienza" and Sezione
INFN  ROMA1, \\ Piazzale Aldo Moro 2, I-00185 Roma, Italy \\ $^2$
Departament d'Astronomia i Astrof\'{\i}sica, Universitat de Val\`encia,
46100 Burjassot, Val\`encia, Spain
}

\date{Accepted ???? Month ??.
      Received 2002 Month ??;
      in original form 2002 Month ??}
\pagerange{\pageref{firstpage}--\pageref{lastpage}}
\pubyear{2002}
\begin{document}

\maketitle

\label{firstpage}

\begin{abstract}
In this paper we study the effects of rotation on the frequencies
of the quasi-normal modes of a proto-neutron star (PNS) born in a gravitational 
collapse during the first minute of life.
Our analysis starts a few tenths of seconds 
after the PNS formation, when
the stellar evolution can be described by a sequence of equilibrium
configurations.
We use the evolutionary models  developed by Pons et al. (1999; 2001)
that describe how a non rotating star cools down and contracts while 
neutrino diffusion and thermalization processes dominate the stellar dynamics.
For assigned values of the evolution time,
we set the star into slow rotation and integrate the equations of stellar 
perturbations in the Cowling approximation, both in the time domain and in
the frequency domain, to find  the quasi-normal mode frequencies.
We study the secular instability of the g-modes, that are present in the
oscillation spectrum due to the intense entropy and composition  
gradients that develop in the stellar interior, and  we provide an estimate  of the 
growth time of the unstable modes based on a  post-Newtonian formula.
\end{abstract}


\section{INTRODUCTION}

It is well known that the frequencies of quasi normal modes (QNMs) of a star 
depend on its internal structure, and therefore on the particular evolutionary
phase the star is going through. 
In a recent paper \cite{FMP}, to be referred to hereafter as FMP,
it has been shown that these frequencies  change 
during the first minute of life of a proto-neutron star
(PNS) born in a gravitational collapse, and that
the changes are mainly due to  neutrino diffusion and thermalisation
processes which smooth out the high entropy gradients that develop
in the stellar interior. 
The models  used in FMP were developed in Pons et al. (1999;2001) and describe the
stellar evolution in terms of a sequence of equilibrium configurations; this
`quasi-stationary' approach has been shown to become appropriate
a few tenths of seconds after the bounce following stellar core collapse
which  gives birth to the PNS.
The study carried out in FMP shows that
the frequency of all QNMs of a newly born PNS are much smaller than 
those of the cold NS which forms at the end of the  evolution.
Indeed, they initially cluster in a narrow region  ($\nu \in [600,1500]$ Hz) and 
begin to differentiate after less than a second.
Unlike in zero temperature, chemically homogeneous stars,
the frequency of the fundamental mode (\f-mode) of a PNS
does not scale as the square root of the average density,
though the star is cooling and contracting;
in addition,  due to the strong thermal gradients
that caracterize the initial life of the PNS, gravity modes (\g-modes)
are also present in the oscillation spectrum, and their frequencies are 
much higher than those of the core \g-modes of cold NS's \cite{reisen,Lai}.

In FMP all stellar models  were assumed to be non rotating; this choice was 
motivated by the need of isolating the effects of thermal and chemical evolution
on the QNM spectrum. However, NSs are expected to be born with a significant
amount of angular momentum, and the aim of this paper is to 
investigate how rotation modifies the  picture described above.
In particular, we shall explore the possibility that  the \g-modes 
become unstable due to Chandrasekhar-Friedman-Schutz (CFS) instability.
This  instability was first discovered by Chandrasekhar (1970) 
for the $m=2$ bar
mode of an incompressible Maclaurin spheroid, and  was later
shown to act in every rotating star by Friedman and Schutz  (1978a,b).

The CFS instability of the fundamental mode (\f-mode)
has been studied extensively in the literature  in the framework of
the newtonian theory of stellar perturbations
\cite{newt1,newt5},  later generalized to the post-newtonian approximation
\cite{cutlin}, and more recently
for fully relativistic, fastly rotating stars \cite{yo1,yo3}.
These studies show that the \f-mode instability acts
at very high stellar rotation rates, comparable to the break-up velocity limit 
of the star. It was also found that, unless the temperature
is very low, viscous dissipation mechanisms tend to 
stabilise the \f-mode  instability.
It should be stressed that, except that in Morsink, Stergioulas \& Blattning
(1999) where a more 
realistic equation of state (EOS) has been considered, all these studies 
use a one parameter, polytropic EOS.
The \r-mode instability has  also extensively been studied in recent years after 
Andersson (1998) pointed out that it is generic for every rotating star,
and since the coupling of the \r-mode  with the current multipoles is strong, 
it was proposed that this instability plays an important  role in nascent 
neutron stars.
In principle, all quasinormal modes  of a rotating star
are unstable for appropriate values of the stellar angular 
velocity $\Omega$, and of the spherical harmonic index $m$, but 
the more  interesting ones are those for which the instability
sets in for a small value of $\Omega$.
In this respect, it should be noted that since in the no rotation limit
the  \g-modes have frequencies  lower than the \f-mode frequency, 
they may become unstable for relatively small values of the angular velocity.
The \g-mode instability has been studied only for zero temperature
stars using the newtonian theory of stellar perturbations in the 
Cowling approximation. This study regarded  a particular class of \g-modes,
for which the buoyancy is provided by the gradient of proton to neutron ratio
in the interior of the star \cite{Lai}.  
In this paper we study the onset of the CFS instability of 
the lowest \g-modes  of a newly born, hot proto-neutron star using, 
as mentioned above, evolutionary models that
take into account the physical processes occurring in the early life of the star.
We use the relativistic theory of stellar perturbations for a
slowly rotating star in the Cowling approximation,
which is known to reproduce with a good accuracy the  \g-mode frequencies
because the gravitational perturbation
induced by \g-modes is much smaller than that associated to the \f-mode. 

The  paper is structured as follows. In section 2 we write
the equations that describe  a perturbed,
slowly rotating, relativistic star both in the time and in the frequency domain
in the Cowling approximation. 
In section 3  we present and discuss the numerical results  both for the
fundamental mode and for the lowest \g-modes; we estimate the growth time 
of the unstable \g-modes  using post-newtonian formulae
and draw the conclusions of our study.

\section{FORMULATION OF THE PROBLEM}

\subsection{The background model}

We consider a relativistic star in uniform rotation  with an angular
velocity $\Omega$ so slow that
the distortion of its figure  from spherical symmetry is
of order $\Omega^2$, and can be ignored.
We expand all equations
with respect the parameter $\veps = \Omega / \Omega_K$, 
where $\Omega_K=\sqrt{\frac{M}{R^3}}$, and retain only first order terms
${\cal O}( \veps )$.
On these assumptions, the metric can be written as \cite{hartle}
\begin{equation}
  \label{metric}
  ds^2 = -e^{2\nu(r)}dt^2
  + e^{2\lambda(r)} dr^2 + r^2\( d\theta^2  + \sin^2\theta d\phi^2\)
  - 2\veps \omega(r)\sin^2\theta dt d\phi.
\end{equation}
The function $\omega(r)$ satisfies a second order linear equation 
\beq
\label{rot2}
\varpi_{,r,r}+\frac{4}{r}\varpi_{,r}-(\lambda+\nu )_{,r}\left(\varpi_{,r}+
\frac{4}{r}\varpi\right)=0,\\
\eeq
where we have defined
\beq
\label{rot3}
\varpi =\Omega -\omega(r).\\
\eeq
In the vacuum outside the star,  $\lambda$ and $\nu$ reduce to the Schwarzschild
functions, and the solution of  eq. (\ref{rot2}) can be written as
\beq
\label{rot5}
\varpi=\Omega-2Jr^{-3},\\
\eeq
where $J$ is the angular momentum of the star.
The star is assumed to be composed by a perfect fluid, whose 
energy momentum tensor is
\begin{equation}
  \label{en_mom_tensor}
  T_{\mu\nu} = \(p + \rho\) u_{\mu}u_{\nu} + pg_{\mu\nu},
\end{equation}
with pressure $p$, energy density $\rho$ and four-velocity components
$  u^{\mu} = [ e^{-\nu},0,0,\Omega e^{-\nu} ]$.
The metric functions $\nu(r),\lambda(r)$ are found by solving Einstein's
equations for a spherically symmetric, non rotating star, 
which couple the metric components to the fluid variables.
As  already mentioned in the introduction we shall  use as a background the models  of 
evolving proto-neutron stars developed in Pons et al. (1999;2001) and used
in FMP; we shall choose different values of the evolution time starting from
$t =  0.5~$s after the formation of the proto-neutron star,
when the quasi--stationary description is appropriate
to represent the stellar evolution.

\subsection{The perturbed equations in the Cowling approximation}

In this section we shall briefly outline the equations that describe the
perturbations of a slowly rotating star up to first order in the rotation
parameter $\veps$.
We shall write these equations both in the time domain and in the frequency
domain, because we have used both approaches to find the mode frequencies.
In both cases we shall assume the oscillations to be adiabatic, so that the
relation
between the Eulerian perturbation of the pressure, $\delta p,$ and of the energy
density,  $\delta\rho,$ is given by
\begin{equation}
  \label{adcond}
    \delta p = \frac{\Gamma_1p}{p + \rho}\delta\rho
      + p'\xi^r\(\frac{\Gamma_1}{\Gamma} - 1\),
\end{equation}
where $\xi^r$ is the radial component of the Lagrangian
displacement $\xi^\mu$, and $\Gamma_1$  and $\Gamma$ are 
\beq
\Gamma_1=\frac{\rho+p}{p}\left(\frac{\partial p}{\partial \rho}
\right)_{s,Y_L},\qquad
\Gamma= \frac{\rho+p}{p}~\frac{p'}{\rho'}~;
\eeq
a prime indicates differentiation with respect to r, and $Y_L$ is the
lepton fraction.
In the following equations we shall also  use  the speed of sound $C_s$ given by 
\begin{equation}
  C^2_s = \left(\frac{\partial p}{\partial \rho} \right)_{s,Y_L}= 
  \frac{\Gamma_1}{\Gamma}\frac{p'}{\rho'}.
  \end{equation}

The complete set of the perturbed Einstein equations has been derived
using the BCL gauge (Battiston, Cazzola \& Lucaroni 1971) in Ruoff,
Stavridis \& Kokkotas (2002). The Cowling limit of these equations 
was studied in Ruoff, Stavridis \& Kokkotas (2003), (RSK) for polytropic
relativistic equations of state. 
In this work we will use the Cowling approximation,
i.e. we shall neglect the contribution of the gravitational perturbations.
In this approach, we need to consider only the fluid perturbations, 
i.e.  the three components of the velocity perturbations $\delta u_i$,
the perturbation of the energy density $\delta\rho$
and the radial component of the displacement vector $\xi^\mu$. The perturbation
of the pressure is related to $\delta\rho$ through the adiabatic condition
(\ref{adcond}).
We expand the fluid perturbations  as
\begin{eqnarray}
  \label{dur}
  \delta u_r &=& -e^{\nu} \sum_{l,m} u_1^{lm} Y_{lm},\\\nonumber
%
%  \label{duth}
  \delta u_{\theta} &=& -e^{\nu} \sum_{l,m} \left[
    \tilde u_2^{lm} \d_{\theta} Y_{lm}
    - \tilde u_3^{lm} {\d_{\phi} Y_{lm} \over \sin\theta } \right],\\\nonumber
%
%  \label{duphi}
  \delta u_{\phi} &=&  -e^{\nu}
  \sum_{l,m} \left[ \tilde u_2^{lm} \d_{\phi} Y_{lm}
    + \tilde u_3^{lm} \sin\theta\d_{\theta} Y_{lm} \right],\\\nonumber
%
%  \label{dutet}
  \delta \rho &=& \sum_{l,m}{\frac{\(p + \rho\)^2}{\Gamma_1p}
    \(H^{lm} - \xi^{lm}\)Y_{lm}},\\\nonumber
%
%  \label{ducsi}
  \xi^r &=& \bigg[\nu'\(1 - \frac{\Gamma_1}{\Gamma}\)\bigg]^{-1}
  \sum_{l,m}{\xi^{lm} Y_{lm}},\nonumber
\end{eqnarray}
where $ H = \delta p/(p+\rho)$ is the enthalpy.

\subsubsection{Perturbed equations in the time domain}
The final set of perturbed equations in the time domain is 
\begin{eqnarray}
  \label{dtHb}
  \nonumber
  \(\d_t + \I m\Omega\)H &=& e^{2\nu-2\lam}\bigg\{C_s^2\bigg[u_1' +
  \(2\nu' - \lam' + \frac{2}{r}\)u_1 - e^{2\lam}\frac{\Lambda}{r^2}u_2
  + 2\I m\varpi e^{2\lam-2\nu}H\bigg] - \nu'u_1\bigg\},\\\nonumber
%
%  \label{dtu1b}
  \(\d_t + \I m\Omega\)u_1 &=& H'
  + \frac{p'}{\Gamma_1 p}\bigg[\(\frac{\Gamma_1}{\Gamma} - 1\)H + \xi\bigg]
  - B\(\I mu_2 + \cLa u_3\),\\
%
%  \label{dtu2b}
  \(\d_t + \I m\Omega\)u_2 &=& H
  + 2\frac{\varpi}{\Lambda}\(\I mu_2 + \cLc u_3\)
  - \frac{\I mr^2}{\Lambda}e^{-2\lam}Bu_1,\\\nonumber
%
%  \label{dtu3b}
  \(\d_t + \I m\Omega\)u_3 &=&
  2\frac{\varpi}{\Lambda}\(\I m u_3 - \cLc u_2\)
  + \frac{r^2}{\Lambda}e^{-2\lam}B\cLb u_1,\\\nonumber
%
%  \label{dtxib}
  \(\d_t + \I m\Omega\)\xi &=& \nu'\(\frac{\Gamma_1}{\Gamma} - 1\)
  e^{2\nu-2\lam}u_1.\nonumber
\end{eqnarray}
where:
\begin{eqnarray}
  \Lambda &=& l(l+1),\qquad\quad
    Q_{lm} := \sqrt{\frac{(l-m)(l+m)}{(2l-1)(2l+1)}},\\\nonumber
B &=& \omega' + 2\varpi\(\nu' - \frac{1}{r}\)  .  
\end{eqnarray}
In eqs. (\ref{dtHb}) we have omitted the indices $lm$ in the perturbed variables
and we have introduced the new   quantities
\begin{eqnarray}
\nonumber
u_2 &:=& \tilde{u}_2 + \I m {\varpi r^2 \over \Lambda} e^{-2\nu} H , \\
\nonumber
u_3 &:=& \tilde{u}_3 -  {\varpi r^2 \over \Lambda} e^{-2\nu} \cLb H . 
\nonumber
\end{eqnarray}
The operators $\cLa$, $\cLb$, and $\cLc$ are the same as in RSK 
and are defined by their action on a perturbation variable $P^{lm}$
\begin{eqnarray}
  \label{cLa}
  \cLa P^{lm} &=& (l-1)Q_{lm}P^{l-1m} - (l+2)Q_{l+1m}P^{l+1m},\\\nonumber
  \label{cLb}
  \cLb P^{lm} &=& -(l+1)Q_{lm}P^{l-1m} + lQ_{l+1m}P^{l+1m},\\\nonumber
  \label{cLc}
  \cLc P^{lm} &=& (l-1)(l+1)Q_{lm}P^{l-1m} + l(l+2)Q_{l+1m}P^{l+1m}.\nonumber
\end{eqnarray}
We must stress here that the above system is an infinite coupled system of
differential equations ranging from $l=m$ to $l=\infty$ 
and that, in order to solve it numerically, we need to truncate 
the couplings  to a finite value of $l_{\rm max}$.

In this work we want to study how the \f and \g modes,
that have polar parity in the non rotating limit,
are affected by rotation.
When the star rotates, the equations that  describe a 
polar perturbation with  harmonic index $l$
acquire rotational corrections  with polar parity and index $l$,
(for instance the terms $\I m\Omega H$ 
and $2\I m\varpi e^{2\lam-2\nu}H$ in the first of eqs. (\ref{dtHb}))
and with  axial parity and index  $l\pm 1$ 
(for instance the term $\cLa u_3$  in the second of eqs. (\ref{dtHb})).

Since it has been shown that the polar rotational corrections are dominant 
for the modes under investigation \cite{kojima97}, in our study we  shall neglect the 
$\pm 1$ axial  rotational corrections.
Thus, the modes we are considering are those indicated in \cite{LFA1}
as the ``polar led hybrid modes".
We choose $l_{\rm max}=2$ and $m=2$. We have checked that considering higher values
of $l_{\rm max}$ would change the results by less than $1\%$.

Eqs. (\ref{dtHb}) have been numerically integrated
giving  an initial gaussian pulse at the enthalpy variable $H$. 
The frequencies  of the modes are identified by looking at the peaks of 
the fast fourier transfom (FFT) of the outcoming signal.

\subsubsection{Perturbed equations in the frequency domain}
By replacing  in Eqs. (\ref{dtHb}) all time derivatives by $\I\sigma$ and letting 
$H \rightarrow \I H$, $u_3 \rightarrow \I u_3$ and $\xi^r \rightarrow \I \xi^r$,
we easily obtain the real valued set of equations describing the 
eigenvalue problem in the frequency domain. 
We have two ODEs for $H$ and $u_1$ and three
algebraic relations for $u_2$, $u_3$ and $\xi^r$. The relation for
$\xi^r$, which follows from the last of eqs.~(\ref{dtHb}) 
is particularly simple and can
be used  to eliminate that variable from the system. The result is
\begin{eqnarray}
  \label{drH}
  \nonumber
  H' &=& \(\sigma + m\Omega\)u_1
  - \frac{p'}{\Gamma_1 p}\(\frac{\Gamma_1}{\Gamma} - 1\)\bigg[H
  -\(\sigma + m\Omega\)^{-1}\nu'e^{2\nu-2\lam}u_1\bigg]
  + B\(mu_2 + \cLa u_3\),\\\nonumber
%
%  \label{dru1}
  u_1' &=& -\(2\nu' - \lam' + \frac{2}{r}\)u_1 + 2 m\varpi e^{2\lam-2\nu}H
  + e^{2\lam}\frac{\Lambda}{r^2}u_2
  - C_s^{-2}\bigg[\(\sigma + m\Omega\)e^{2\lam-2\nu} H - \nu'u_1\bigg],\\
%
%  \label{u2alg}
  u_2 &=& \Sigma^{-1}\bigg[H - \frac{mr^2}{\Lambda}e^{-2\lam} Bu_1
  + \frac{2\varpi}{\Lambda}\cLc u_3\bigg],\\
  \nonumber
%
%  \label{u3alg}
  u_3 &=& -\Sigma^{-1}\bigg[\frac{r^2}{\Lambda}e^{-2\lam}B\cLb u_1
  - \frac{2\varpi}{\Lambda}\cLc u_2\bigg],
  \nonumber
\end{eqnarray}
where we have defined 
\begin{eqnarray}
  \Sigma := \sigma + m\Omega - \frac{2m\varpi}{\Lambda}.
\end{eqnarray}
An inspection of eqs. (\ref{drH}) shows that they become singular when
$\Sigma :=0.$  For any assigned value of $l,m,\Omega$ and
$\sigma$, this may happen inside the star
in a certain domain of the radial coordinate $r$ which would depend on
the values of the function $\varpi$.  This occurrence would generate the
so-called continuous spectrum.
As explained in RSK, for fixed values of the 
compactness of the star $M/R$, the frequency region were the continuous spectrum
extends pratically depends on the values of $\varpi$ 
at the center and at the surface of the star and on the number of 
maximum couplings $l_{\rm max}$ that is considered.
By following the procedure explained in RSK (section 2.3), it is easy to show that
in the  non-axisymmetric case ($m\neq0$) and for $l_{\rm max}=2$
the continuous spectrum extends to the following region 
\beq
\label{contspec}
2 \left( {\varpi_c \over 3} - \Omega \right) 
\le \sigma \le 
2 \left( {\varpi_s \over 3} - \Omega \right).
\eeq

Assuming $l=m=2$,  Eqs. (\ref{drH})
can be written in the following simplified form
\beqar
\label{drHl2final}
H' &=& \bigg[ {m B \over \Sigma} 
-{p' \over \Gamma_1 p} \( {\Gamma_1 \over \Gamma} - 1 \)  \bigg] H
      + \bigg[ \sigma + m \Omega + ( \sigma + m\Omega )^{-1} 
{p' \over \Gamma_1 p} \( {\Gamma_1 \over \Gamma} - 1 \) 
\nu' e^{2\nu-2\lambda} \bigg] u_1 \\
\nonumber
%\label{dru1l2final}
   u_1' &=& \bigg[ -2\nu' + \lambda' - {2 \over r} + C_s^{-2} \nu' - {m B \over \Sigma} \bigg] u_1 
            + \bigg[ e^{2\lambda-2\nu} \bigg( 2 m \varpi  - C_s^{-2}  ( \sigma + m\Omega ) \bigg) 
            + {\Lambda e^{2\lambda} \over r^2 \Sigma}  \bigg] H \
\eeqar
where we have used the third of eqs. (\ref{drH}) to eliminate
$u_2.$ 
To find the mode frequencies, we integrate these 
two ODEs  by imposing that the variables have a regular 
behaviour near the center, i.e.
\begin{equation}
H \sim r^l, \qquad\quad 
u_1 \sim  r^{l-1},
\end{equation}
and select those frequencies for which 
the Lagrangian perturbation of the pressure vanishes at the surface, i.e.
\beq
\Delta p = ( \sigma + m \Omega ) H(R) - \nu'(R) e^{4\nu(R)} u_1(R) = 0.
\eeq

\section{Results}
In order to find the frequency of the quasi normal modes,
we have numerically integrated the perturbed equations both in the time
and in the frequency domain for different values of the evolution time $t_{ev}$,
and for selected values of the rotation parameter $\veps.$
We choose the  rotation rate  to vary within  $0\leq \veps \leq 0.4$ because 
from preliminary calculations  we find that for the models under consideration
the mass shedding limit does not exceed $\veps=0.4-0.5$. 

We consider the evolutionary model labelled as model A in FMP, in which
the equation of state of baryonic matter is a finite-temperature, 
field-theoretical model solved at the mean field level.  
Electrons and muons are included in the 
models as non interacting particles, being the contribution due to their 
interactions much smaller than that of the free Fermi gas, and  
neutrino transport is treated using  the diffusion approximation.
The evolution time interval we consider covers the first minute of 
life of the proto-neutron star, from  $ t_{ev} = 0.5~ s$ 
to $ t_{ev} =  40 ~s$, when processes related to neutrino
diffusion and thermalization become negligible.
The gravitational mass of the star at $t_{ev}=0.2~s$ is
$M=1.56~M_\odot$, and at $t_{ev}=40~s$ becomes
$M=1.46~M_\odot$. The difference in
gravitational mass  between the initial and final configuration
is radiated away by neutrinos during the PNS evolution.
The radius of the initial configuration is $R=23.7$ km and reduces to $R=12.8$ km
at $t_{ev}=40~s$.

For $t_{ev} \lappreq 20~s$ the stellar models  are convectively unstable,
and the code which integrates the perturbed equations in the time domain  
explodes after some time, which is too short  to accurately calculate
the low frequencies of the \g-modes.
Conversely, the  code which  integrates the equations in the
frequency domain is well behaved even when convective instability is present,
and therefore for $t_{ev} < 20~s$ we  use the frequency domain approach.
After that time both methods can be applied and the results  agree better than $~5\%$.

It is worth stressing here once more that the perturbed equations in the
frequency domain present a singular structure which makes impossible
their numerical integration in the continuous spectrum region.
However, for the stellar models we use  and for the mode frequencies we are
interested in, we find that the continuous spectrum lays in the 
negative frequency range.

\begin{figure}
\begin{center}
\centerline{\mbox{
\epsfig{figure=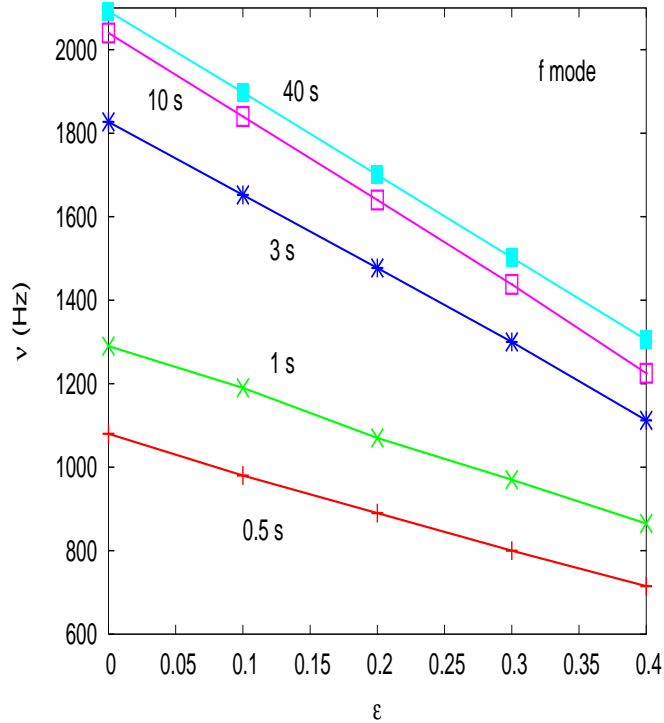,width=10.0cm,height=9.0cm,angle=-90}
}}
\vspace{0.5cm}
\caption{
The frequency of  the fundamental  mode  of the evolving proto-neutron star is
plotted as a function  of the rotational parameter $\veps = \Omega / \Omega_K,$
for assigned values of  the time elapsed from the gravitational collapse.
We see that as the time increases, the frequency increases and tends
to that of the cold neutron star which forms at the end of the evolutionary
process. Reminding that the onset of the CFS instability occurs when the 
mode frequency becomes zero, we see that the \f-mode would
become unstable only for extremely high values of the rotational parameter, 
 as it is for cold stars.
}
\label{fig1}
\end{center}
\end{figure}

\begin{figure}
\begin{center}
\centerline{\mbox{
\epsfig{figure=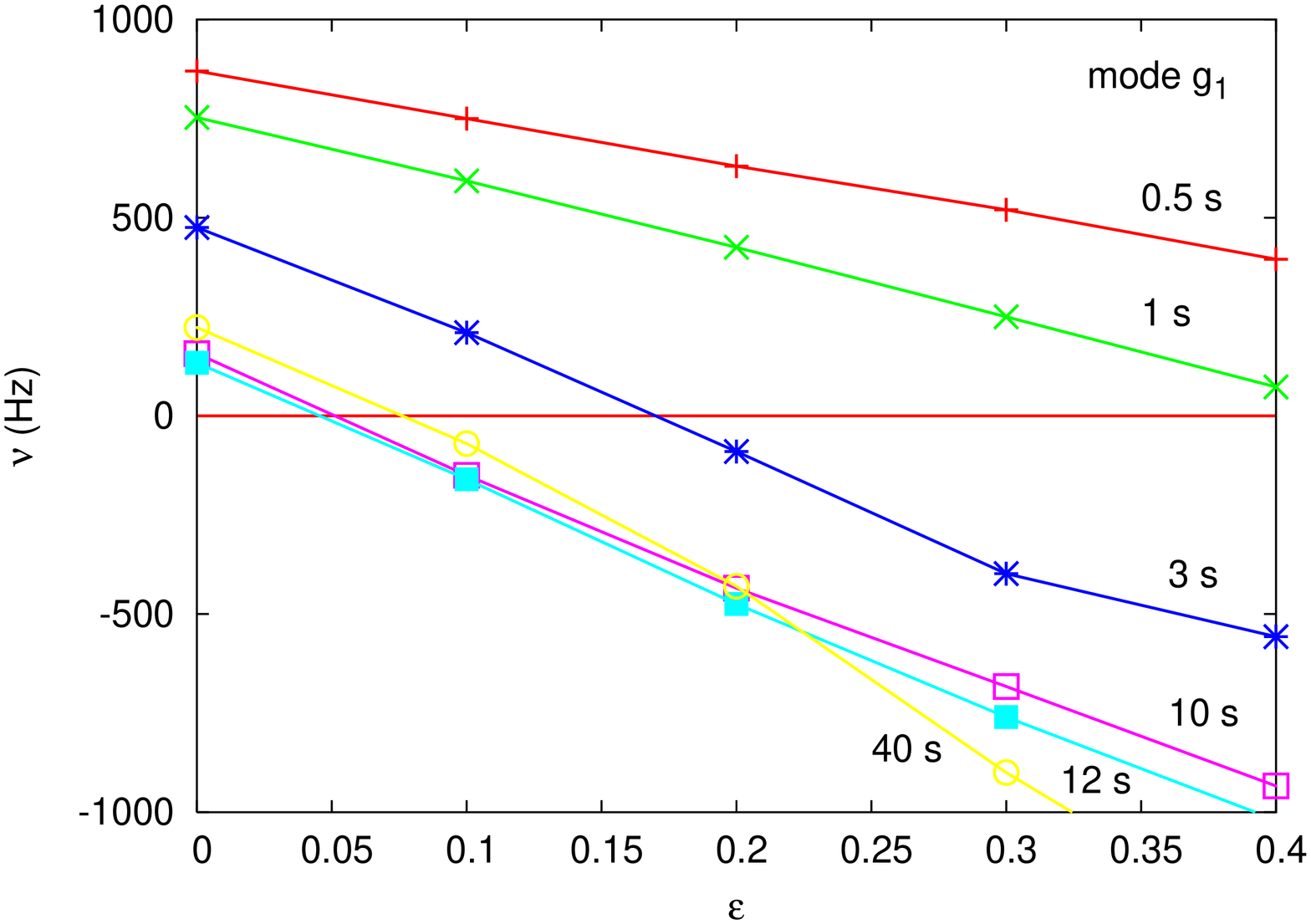,width=10.0cm,height=9.0cm,angle=-90}
\epsfig{figure=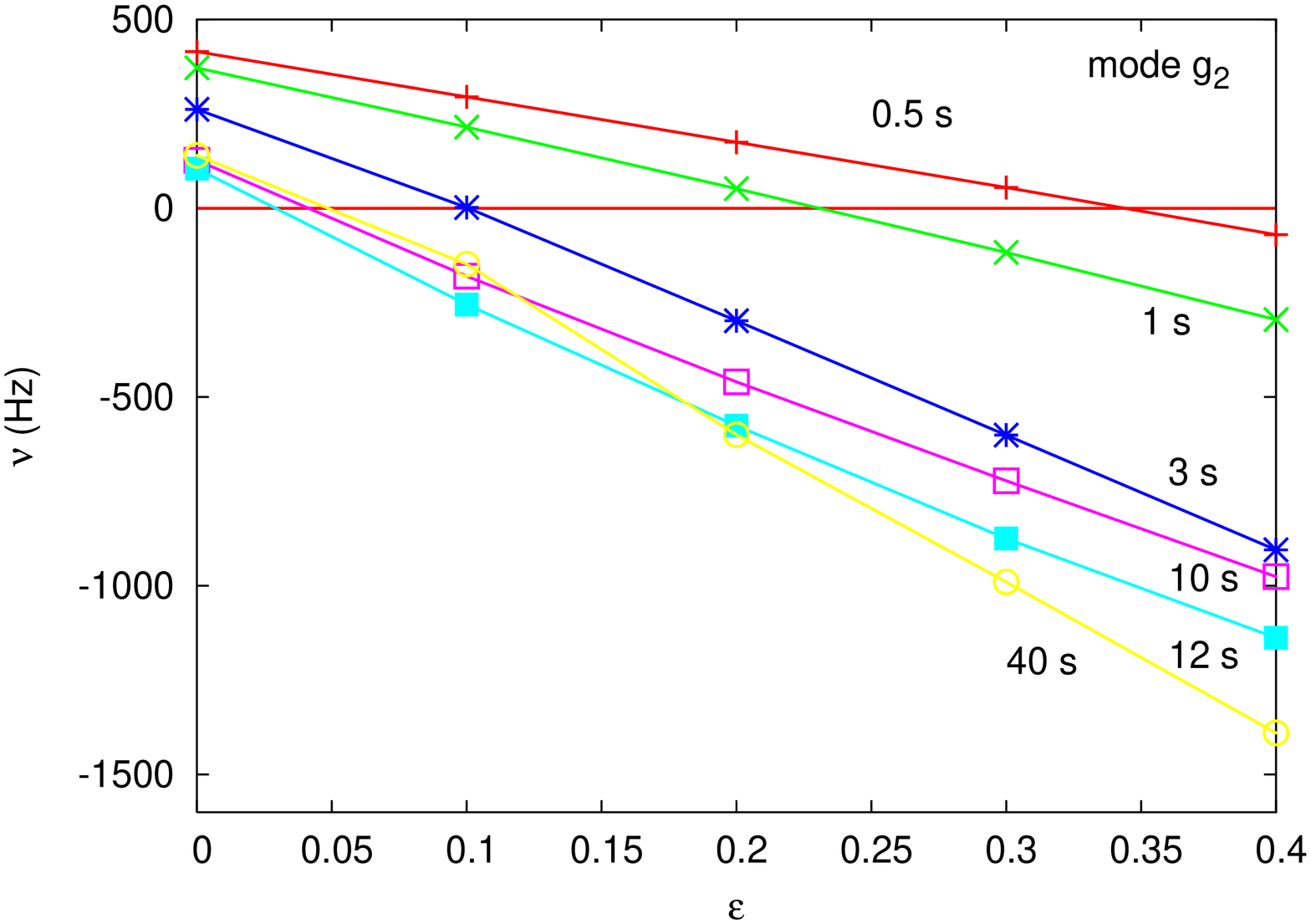,width=10.0cm,height=9.0cm,angle=-90}
}}

\vspace{0.5cm}
\caption{
The frequency of  the \1 and \2  modes  of the proto-neutron star are
plotted, as in figure 1, as functions  of the rotational parameter
for assigned values of  the time elapsed from the gravitational collapse.
Unlike the \f-mode, as the time increases the frequency  of the \g-modes
decreases, reaches a minimum  at about $t_{ev}=12$ and
then slightly increases (see text). 
We see that for both modes  the CFS instability sets in at values of the
rotational parameter much lower that that needed for the \f-mode.
}
\label{fig2}
\end{center}
\end{figure}

The main results of this work are summarized in figures \ref{fig1}
and \ref{fig2}, where we plot the frequencies of the \f, \1 and \2 modes as a 
function of the rotation parameter $\veps= \Omega / \Omega_K$,
for different values of the evolution time in the more interesting phases 
of the cooling process.

It should be reminded that the onset of the CFS instability is signaled by the
vanishing of the  mode frequency for some value of the angular
velocity (neutral point).
From figure \ref{fig1} and \ref{fig2}  we see that while the 
\f-mode does not become  unstable
during the first minute of the PNS life, both the \1 and the \2 modes do become
unstable.
The \1 frequency remains positive during the first second, but 
at later times vanishes  for very low values of $\veps.$ For instance, at
$t_{ev}=3 ~s$ it crosses the zero axis for $\Omega =0.17~\Omega_K,$ even though
its value for the corresponding nonrotating star is still quite high,
$\nu_{g_1}=486 ~$Hz.
The behaviour of the \2 mode is similar, but being the frequency lower the
instability sets in at lower rotation rates.
At later times, the {\bf g}-modes frequencies decrease,
reach a minimum  for $t_{ev}=12 ~s$ and then slightly increase.
This behaviour can be attributed to the fact that
during the first 10-12 seconds the dynamical evolution of the star
is dominated by strong entropy gradients that progressively smooth out. 
After about $12~s$ the entropy has become nearly
constant throughout the star
and {\bf g}-modes due to composition gradients take over.

It should be mentioned that in FMP we studied also a second model of evolving
proto-neutron star, labelled as model B.
The main difference between the
two models is that model A has an EOS softer than  model B, and that at some
point of the evolution a quark core forms in the interior of  model B.
We have integrated the perturbed equations (\ref{dtHb}) and (\ref{drHl2final})
also for model B, finding results entirely similar to those described above 
for model A; this indicates that
 that the quark core that develops at some point of the evolution 
in model B does not affect the overall  properties of the modes in a relevant way.

\subsection{The  growth time of the unstable modes}
A mode  instability is relevant if its growth time is sufficiently small
with respect to the timescales typical of the stellar dynamics,
i.e. if the instability has sufficient time to grow before other processes damp
it out or the structure of the evolving star changes.
In this section we will give an ``order of magnitude" estimate 
of the growth time of the \g-modes
for the hot proto-neutron stars under investigation.
Following Lai (1999), we shall evaluate the  mode energy  in the rotating frame
using the expression 
given in  Friedman \& Schutz (1978a)
\beq
E = {1 \over 2} \int \left[  \rho \; (\sigma +m \Omega)^2 \;
\vec{\xi}^*\; \cdot \;\vec{\xi}+
 \( {\delta p \over \rho} - \delta \Phi \) \delta \rho^* 
+\left( \vec{\nabla} \; \cdot \;\vec{\xi}\right)
\vec{\xi}^*\; \cdot \; 
\left(\vec{\nabla} p -C_s^2 \vec{\nabla}\rho  \right)
\right] d^3 x,
\label{energy}
\eeq
and  compute  the growth time associated to the dissipative process we are
considering, i.e. the gravitational radiation reaction,  using the expression
\beq
{1 \over \tau_{gr}} = - {1 \over 2E} {dE \over dt}. 
\eeq  
It is useful to remind that the relation between the Lagrangian
displacement and the four-velocity of a perturbed fluid element is 
\beq
\delta u^k=\I (\sigma+m\Omega)e^{-\nu} ~\xi^k, \qquad\quad k=1,3.
\eeq
Since we are working in the Cowling approximation, the term
$\delta \Phi$ in eq. (\ref{energy}) will be neglected.
The energy loss due to gravitational waves can be calculated from the 
multipole radiation formula of Lindblom et. al. (1998) given by
\beq
\( {dE \over dt} \)_{\rm gr}= -\sigma \( \sigma + m \Omega \) 
\sum_{l \ge 2} N_l \sigma^{2l} \( |\delta D_{lm}|^2 + |\delta J_{lm}|^2 \),
\label{lum}
\eeq
where the coupling constant $N_l$ is given by
\beq
N_l = { 4 \pi G \over c^{2l+1} } { (l+1)(l+2) \over l(l-1) [(2l+1)!!]^2 },
\eeq
and $\delta D_{lm}$ and $\delta J_{lm}$ are the mass and current 
multipoles  of the perturbed fluid.
The current multipoles $\delta J_{lm}$ are
associated with the axial spherical harmonics;  since we are interested in the 
{\bf g}-modes, for which the effect of the  coupling between polar
and axial perturbations is negligible (indeed  we did not include it 
in the perturbed equations),  we shall neglect the $\delta J_{lm}$
contribution to the gravitational luminosity.
The mass multipoles can be evaluated from the following integral expression
\beq
\delta D_{lm} = \int { r^l \, \delta \rho \, Y_{lm}^* d^3 x }.
\label{dlm}
\eeq
The  growth times of the unstable \1 modes shown in figure \ref{fig2}
are summarized in Table 1. From these results we see that
the growth time appears to be orders of magnitude larger than
the timescale on which the star evolves, which is of the order of tens of
seconds.  Although we are aware that  the
estimate based on the newtonian expressions (\ref{energy}) and (\ref{dlm}) 
is a quite crude one, the growth time  is so much larger than
the evolutionary timescale  that it is reasonable to conclude that 
the CFS instability of the lowest \g-mode is unlikely to
play any relevant role in the early evolution of proto-neutron stars.
Similar conclusions can be drawn for the fundamental mode and
for higher order {\bf g}-modes.

\section{Aknowledgements}

We would like to thank G. Miniutti for suggesting that the \g-modes of newly born
PNSs may be subjected to the CFS instability, and N. Andersson for useful
discussions on the subjects related to this paper.

This work has been supported by the EU Program 'Improving the Human Research 
Potential and the Socio-Economic Knowledge Base' 
( Research Training Network Contract HPRN-CT-2000-00137 ).

\label{lastpage}
\newpage

%%%%%%%%%%%%%%%%%%%%%%%%%%%%%%%%%%%%%%%%%%%%%%%%%%%%%%%%%%%%%%%
% Table 2
\begin{table}
\centering
\caption{Growth times for unstable $g_1$ mode of Model A for $t_{ev}=3~s$, 
$t_{ev}=10~s$, $t_{ev}=12~s$, and $t_{ev}=40~s$}
\begin{tabular}{*{3}{c}}
\hline\hline
$\veps$ & $\nu$~(Hz) & $\tau_{gr}~ (s) $ \\[0.5ex]
\hline\hline
0.1 & 200 & \dots \\[0.5ex]
0.2 & -90 & -1.5 $10^{9}$  \\[0.5ex]
0.3 & -683 & -2 $10^6$  \\[0.5ex]
0.4 & -789 & -2.2 $10^{4}$  \\[0.5ex]
\hline\hline
\label{modelAt3}
\end{tabular}
\begin{tabular}{*{3}{c}}
\hline\hline
$\veps$ & $\nu$~(Hz) & $\tau_{gr}~ (s) $ \\[0.5ex]
\hline\hline
0.1 & -150 & -8.3 $10^{9}$  \\[0.5ex]
0.2 & -435 & -6.5 $10^{6}$  \\[0.5ex]
0.3 & -683 & -4 $10^5$  \\[0.5ex]
0.4 & -910 & -2.4 $10^{3}$  \\[0.5ex]
\hline\hline
\label{modelAt10}
\end{tabular}
\begin{tabular}{*{3}{c}}
\hline\hline
$\veps$ & $\nu$~(Hz) & $\tau_{gr}~ (s) $ \\[0.5ex]
\hline\hline
0.1 & -160 & -4.6 $10^{7}$  \\[0.5ex]
0.2 & -475 & -1.7 $10^{6}$  \\[0.5ex]
0.3 & -760 & -3.4 $10^5$  \\[0.5ex]
0.4 & -1020 & -2.7 $10^{5}$  \\[0.5ex]
\hline\hline
\label{modelAt12}
\end{tabular}
\begin{tabular}{*{3}{c}}
\hline\hline
$\veps$ & $\nu$~(Hz) & $\tau_{gr}~ (s) $ \\[0.5ex]
\hline\hline
0.1 & -70 & -3.7 $10^{10}$  \\[0.5ex]
0.2 & -430 & -1.3  $10^{9}$ \\[0.5ex]
0.3 & -900 & -3.6 $10^6$  \\[0.5ex]
0.4 & -1250 & -1.3 $10^5$  \\[0.5ex]
\hline\hline
\label{modelAt40}
\end{tabular}
\end{table}
%%%%%%%%%%%%%%%%%%%%%%%%%%%%%%%%%%%%%%%%%%%%%%%%%%%%%%%%%%%%%%%%%%%%%%%%%%%
%%%%%%%%%%%%%%%%%%%%%%%%%%%%%%%%%%%%%%%%%%%%%%%%%%%%%%%%%%%%%%%%%%%%%%%%%%%
%%%%%%%%%%%%%%%%%%%%%%%%%  FIGURES  %%%%%%%%%%%%%%%%%%%%%%%%%%%%%%%%%%%%%
%%%%%%%%%%%%%%%%%%%%%%%%%%%%%%%%%%%%%%%%%%%%%%%%%%%%%%%%%%%%%%%
%%  FIG1 %%

%%%%%%%%%%%%%%%%%%%%%%%%%%%%%%%%%%%%%%%%%%%%%%%%%%%%%%%%%%%%%%%

\end{document}